\documentclass[12pt,preprint]{aastex}
\usepackage{rotating}
\usepackage{epsfig}
\slugcomment{Submitted to Astrophysical Journal (Letters)}
\shorttitle{A Highly Collimated, Young and High Velocity CO(2-1) Outflow in OMC1 South}
\shortauthors{Zapata et al.}

\begin{document}

\title{A Highly Collimated, Young and Fast CO(2-1) Outflow in OMC1 South}
\author{Luis A. Zapata,$^{1,2}$ 
Luis F. Rodr\'\i guez\footnote{Centro de Radioastronom\'\i a y Astrof\'\i sica, 
UNAM, Apdo. Postal 3-72 (Xangari), 58089 Morelia, Michoac\'an, M\'exico}, Paul T. P. Ho$^{2}$, 
 Qizhou Zhang$^{2}$, Chunhua Qi\footnote{Harvard-Smithsonian Center for Astrophysics, 
 60 Garden Street, Cambridge, MA 02138, USA}, \\ 
 S. E. Kurtz$^{1}$}

\email{lzapata@astrosmo.unam.mx}
 
\begin{abstract}
We present high angular resolution ($\sim$ 1$''$), sensitive CO(2-1) line observations 
of the region OMC1 South in the Orion Nebula made using the Submillimeter Array (SMA). 
We detect the CO(2-1) high velocity outflow that was first found  
by Rodr\'\i guez-Franco et al. (1999a) with the IRAM 30 m. 
Our observations resolve the outflow, whose velocity-integrated emission has a deconvolved width of 
0\rlap.{''}89 $\pm$ 0\rlap.{''}06 (390 AU) and a projected length of $\sim$ 48${''}$ (21,000 AU)
with very high redshifted and blueshifted gas with velocities of about $\pm$80 km s$^{-1}$. 
This outflow is among the most collimated ($\sim$ 3$^{\circ}$) and youngest  
outflows (600 yr) that have been reported. 
The data show that this collimated outflow has been blowing in the same  
direction during the last 600 yr. 
At high velocities, the CO(2-1) outflow traces an extremely collimated jet, while at
lower velocities the CO emission traces an envelope possibly 
produced by entrainment of ambient gas. 
Furthermore, we also detect for the first time a 
millimeter wavelength continuum source possibly associated with a class I protostar 
that we suggest could be the possible exciting source for this collimated outflow.
However, the bolometric luminosity of this source appears to be far too low
to account for the powerful molecular outflow. 
\vspace{1cm}
\end{abstract}  


\section{Introduction}

The OMC1-S (Orion Molecular Cloud 1 South) region is a young,
compact (40${''}$$\times$40${''}$) and highly active star formation 
zone located about 100$''$ south 
of the Orion-KL region. At least three powerful molecular 
outflows have been suggested to emanate from here. The first is a low velocity (10 km s$^{-1}$)  
bipolar SiO(5-4) outflow with a length of $\sim$ 30${''}$ in the NE(blueshift)-SW(redshift) 
orientation, and that is centered at the position R.A.[J2000]=05:\\ 35:12.8 and Dec.[J2000]=-05:24:11 
(Ziurys, Wilson \& Mauersberger 1990). A second quite extended ($\sim$ 3${'}$), collimated, low 
velocity (5 km s$^{-1}$) bipolar CO outflow, oriented 
NE (blueshift)-SW (redshift) has been reported by Schmid-Burgk et al. (1990), 
and its center is associated with the continuum radio 
source 134-411 (Zapata et al. 2004a, 2004b).      
Finally, there is a high velocity bipolar CO outflow, with a length of 0.07 pc (0.5${'}$) and
 velocities of $-$140 km s$^{-1}$ towards the NW and 88 km s$^{-1}$ 
to the SE (Rodr\'\i guez-Franco et al. 1999a, 1999b). 
The center of this high velocity outflow was proposed to be 
20$''$  north of the 1.3 mm continuum source 
FIR 4 (Rodr\'\i guez-Franco et al. 1999b). As all these outflows were detected with 
single dish observations of modest angular resolution, 
their relation is not clear.    

There are also several relatively large Herbig-Haro outflows 
of length up to 2${'}$ that appear to emanate from the vicinity of this 
region (HH~202, HH 269, HH 529, HH~203/204, HH~530, HH 625
and possibly HH~528) and that were found by Bally, O'Dell \& McCaughrean (2000) 
and O'Dell \& Doi (2003) at optical wavelengths. 
In particular, the HH 625 object is the first optical 
flow of the Orion Nebula that has been associated with an 
embedded molecular outflow (O'Dell  2003 and O'Dell \& Doi 2003).
This object was first detected by O'Dell \& Wong (1996) and
then they found associated faint features in molecular hydrogen (H$_2$) 
seen in the infrared by Kaifu et al. (2000) 
and Stanke, McCaughrean \& Zinnecker (2002). 
It is oriented toward P.A.=325$\degr$ (blueshifted gas) 
in good agreement with the direction (P.A.= 310$\degr$) of the high velocity 
molecular outflow found by Rodr\'\i guez-Franco et al. (1999b).    

In this Letter, we present new sensitive and high angular resolution ($\sim$ 1${''}$) 
CO J=2$\rightarrow$1 observations of the OMC1-S region that allow us to resolve the CO 
outflow first reported by Rodriguez-Franco et al. (1999a), and to propose a possible 
exciting source.

\section{Observations}

The observations were made with the SMA\footnote{The Submillimeter 
Array (SMA) is a joint project between the Smithsonian 
Astrophysical Observatory and the Academia Sinica Institute 
of Astronomy and Astrophysics, and is funded by the Smithsonian
Institution and the Academia Sinica.} during 2004 
September 2. The array was in its "extended" configuration, which includes
21 independent baselines ranging in projected length from 16 to 180 meters.
The phase reference center of the observations was  
R.A.[J2000]$=$05:35:14 and Dec.[J2000]$=$-05:24:00.  
The frequency was centered on the CO(2-1) line at 230.53 GHz in the upper sideband, 
while the lower sideband was centered at 220.53 GHz.
The SMA digital correlator was configured with a band of 32 channels over 104 MHz, 
which provided 3.25 MHz (4.29  km s$^{-1}$) resolution. 
The zenith opacity measured ($\tau_{230GHz}$) with the NRAO tipping
radiometer located at the Caltech Submillimeter Observatory was $\sim$0.04.
The average system temperature was 250 K. The continuum image was constructed by
averaging the line-free channels in the upper side band.
The phase and amplitude calibrators were quasars 0423-013 and 3C120,
 with density fluxes of 2.343$\pm$0.006 Jy and
 0.563$\pm$0.003 Jy, respectively. The uncertainty in the flux scale is estimated to be 
20\%, based on the SMA monitoring of quasars.
Observations of Callisto provided the absolute scale for the flux density calibration. 
Further technical descriptions of the SMA  and its calibration schemes are
found in Ho, Moran \& Lo (2004).
    
The data were calibrated using the MIR software package.
The calibrated data were imaged and analyzed in 
the standard manner using the MIRIAD and AIPS packages. 
We weighted the data using the ROBUST parameter of INVERT set to 5, 
searching for a maximum sensitivity 
in each continuum and line image. The resulting continuum image 
rms was 11 mJy beam$^{-1}$, at an angular resolution of 
$1\rlap.{''}13 \times 0\rlap.{''}95$ with PA = $-73^\circ$, while 
the line image rms noise was 120 mJy beam$^{-1}$ for each channel.
All images were corrected for the primary beam response. 

\section{Discussion}
\subsection{Continuum Emission} 

In the OMC1-S region we found a total of nine 1.3 mm continuum sources (see Figure 1),
seven of them (134-411, 136-359, 136-335, 137-347, 139-409, 144-357 and 144-351) 
have been detected previously at other wavelengths by 
Ali $\&$ Depoy (1995); Feigelson et al. (2002) ; Gaume et al. (1998); 
Mezger, Zylka \& Wink (1990) ; Mundy et al. (1986); Schulz et al. (2001);  
Muench et al. 2002; Lada et al. 2004; Smith et al. (2004); 
Robberto et al. (2004); and Zapata et al. (2004a, 2004b).
The source 132-413 is first reported here and does not have
counterparts at other wavelengths. The source 137-408 is first reported here, but 
possibly is associated with the FIR4 source.  
To refer to these sources we adopted a position-based 
nomenclature after the convention of O'Dell $\&$ Wen (1994). 
The nature of these continuum sources, and their 7 mm counterparts recently detected 
with the VLA, will be discussed in a future paper (Zapata et al. 2005, in preparation).  
  
With our angular resolution and sensitivity we detect for the first time the  
millimeter counterpart of the source 136-359 (see Figure 1). This source is well centered
on the axis of the collimated CO J=2$\rightarrow$1 outflow as described below, and 
resides at the base of the blueshifted jet. We therefore suggest that 136-359 may be the likely 
exciting source of the bipolar molecular outflow. 
The source is located at the position R.A. [J2000]$=$05:35:13.550 and 
DEC. [J2000]$=$-05:23:59.14, has a deconvolved size  $\leq$ 0.6$''$ and an 
integrated continuum flux at 1.3 mm of 116.2$\pm$9.0 mJy. 

This source has  near- and mid-infrared counterparts (Gaume et al. 1998, Lada et al. 
2000; Lada et al. (2000, 2004); Smith et al. 2004 and  Robberto et al. 2005).
This object is also coincident within 0.1$''$ with a 1.3 cm continuum source 
(136-359; Zapata et al. 2004b) and a water maser with low velocity emission 
reported by Gaume et al. (1998). 

Figure 2 shows the spectral energy distribution (SED) of 136-359 combining
data from the centimeter, millimeter and infrared wavelengths. We fitted the SED 
with two modified blackbody functions of the form B$_{\nu}$(T$_d$)(1-$\exp(-\tau)$), 
where $\tau$ is the dust optical depth, and, B$_{\nu}$(T$_d$) is the Planck function
at the dust temperature T$_d$. The opacity was assumed to vary with frequency as 
$\tau$=$\tau_{0}$$(\nu/\nu_0)^{\beta}$, where $\tau_0$=6.0 $\times$ 10$^{-4}$ and 
$\nu_0$ is the frequency at which the optical depth is unity.
From the fit (dotted line), we derive that the dust component has a temperature 
of 83 K and that ${\beta}$ is equal to 0.13. The temperature  derived 
for the other component is 870 K. This hot component might represent a 
more evolved and hot inner core. We note that this dust temperature is higher 
than the typical dust temperatures ($\sim$30 K) observed in other young protostars 
also associated with collimated outflows (VLA 1613, Andr\'e et al. 1993; L1448-mm,
Bachiller et al. 1990; HH24-mm, Chini et al. 1993). This suggests that the 
dust associated with 136-359 is being strongly heated, possibly either by internal 
or external sources.

The grain opacity spectral index for this source is rather small.
Similar small values have been
found in a few other sources (e.g. Kitamura et al. 2002, and
Williams, Fuller \& Sridharan 2004) 
and are usually attributed to growth and evolution of the dust grains within dense, 
dusty regions. An alternative possibility is that we are detecting not dust
emission, but optically-thick free-free emission. However, if interpreted
this way, the large
flux densities observed would imply significant photoionization and the
presence of an early
B-type star, a result inconsistent with the bolometric luminosity of
the source (note, however, that as will be discussed below, the possibility of 
a luminous star is suggested by the large mechanical power of the outflow). 
Finally, the dust emission itself could be optically
thick. Additional high angular resolution observations of
this source in the far-infrared and submillimeter 
regions may favor one of the possibilities.

From the millimeter wavelength flux, and the parameters obtained from
our fit we estimate a total dust and gas mass
for 136-359 of 0.16 M$_{\odot}$ assuming optically thin dust emission, 
 the Raleigh-Jeans  approximation, a gas-to-dust ratio of 100,  
and a typical dust opacity per gram of $\kappa$ = 0.006[$\nu$/245 GHz]$^{\beta}$
where $\beta$ is 0.13 (see Hildebrand et al. 1983).

We integrated the fit of the SED to 136-359 and obtained a bolometric 
luminosity equal to 8 L$_{\odot}$. We estimate that 136-359 has a ratio of
L$_{bol}$/L$_{1.3mm}$= 1.2 $\times$ 10$^5$ $\geq$ 2.0 $\times$ 10$^4$.
Moreover the source is visible at mid-infrared wavelengths (8.8 $\mu$m).
We propose that 136-359 may be a class I source (see Andr\'e, Ward-Thompson, \& Barsony, 1993).

\subsection{CO J=2$\rightarrow$1 Collimated Outflow}

Figure 1 shows the spatial distribution of the integrated 
CO J=2$\rightarrow$1 emission of the blue (-80 to -26 km s$^{-1}$)
and redshifted (22 to 82 km s$^{-1}$) components of  
the bipolar outflow. Our map clearly resolves the structure of the outflow. 
The red component shows a deconvolved width of 0\rlap.{''}89 $\pm$ 0\rlap.{''}06 and 
a length of 30$''$ (390 AU $\times$ 13,200 AU at the distance of 440 pc), 
while the blue component shows a deconvolved width of 0\rlap.{''}65 $\pm$ 0\rlap.{''}02 
 and a length of 18$''$ (280 AU $\times$ 7900 AU). 
 The P. A. of the outflow is 304.4$^{\circ}$$\pm 0.7$$^{\circ}$,
 and we estimate a collimation angle of $\sim$ 3$^{\circ}$.
Only a handful of outflows with similarly small collimation angles 
have been reported in the literature (HH 211, Gueth \& Guilloteau 1999;
HH 212, Zinnecker, McCaughrean \& Rayner 1998, and Lee et al. 2000; HH111, 
Cernicharo \& Reipurth 1996; IRAS 11590-6452, Bourke et al. 1997; 
VLA1623, Yu \& Chernin 1997; IRAS 04166+2706, Tafalla et al. 2004). 
This outflow is also one of the youngest low-mass outflows known, having 
a dynamical age of $\tau_{d}$ = 0.1pc/160 km sec$^{-1}$ = 600 yrs 
(uncorrected for projection). 

Following Su, Zhang \& Lim (2004), we estimate a mass of 0.52 M$_{\odot}$ for the 
redshifted component of the outflow and a mass of 0.20 M$_{\odot}$ for the blueshifted 
component. We used   
the ratio of the flux densities of $^{12}$CO[J=2$\rightarrow$1] 
and $^{13}$CO[J=2$\rightarrow$1], assuming LTE conditions, an  excitation temperature 
of 80 K, an abundance ratio of $^{13}$CO/H$_{2}$ equal to 8.5 $\times$ 10$^{-5}$ (Frerking,
Langer \& Wilson 1982) and an isotopic ratio [$^{12}$CO]/[$^{13}$CO] = 70.5 (Wilson $\&$ Rood 1994) 
for the appropriate galactocentric distance of Orion. The rest of the derived parameters are shown 
in Table 1. As can be seen in this Table, the large mechanical luminosity of the , 
much larger than the bolometric luminosity of 136-359, strongly suggesting that the true exciting s
ource could be an obscured, high luminosity object.

This outflow is indeed extremely powerful for the low bolometric mass derived for the
proposed exciting star. From our results and those of Rodriguez-Franco et al.
 (1999) we estimate a mechanical force of about 
0.03 M$_{\odot}$ km s$^{-1}$ (years)$^{-1}$.
After the observational correlation for mechanical force as a function of
bolometric luminosity shown in Figure 3 of Beuther et al. (2002b), we find that
a bolometric luminosity of about 10$^4$ solar luminosities is expected
for the exciting source. This value is three orders of magnitude larger than
the bolometric luminosity of 8 solar luminosities derived from the spectrum of
136-359 (see Fig. 2).

We believe that the mechanical force estimate is reliable,
and that the discrepancy could be explained if 136-359 is not
the true exciting source. One possibility is that the far-infrared and 
submillimeter emissions of the source
are much larger than the value interpolated in our fit. This could
be possible if the centimeter 
data has a high contamination of free-free emission. 
If this is the case, possibly this source is more likely associated with a
class 0 source and the total luminosity can be increased to $\sim$ 20L$_{\odot}$.
However, even with this increased bolometric
luminosity this outflow remains as one of most efficient
outflows. 

Another possibility is that 136-359 is not the true exciting source but 
just a line-of sight star, part of the cluster but not the true exciting object. 

The exciting object would be a high luminosity, extremely obscured object.
Its bolometric luminosity would come out in the far-infrared, possibly as an
extended source difficult to detect in the available observations.
Finally, 136-359 may be a post-FU Ori object that reached much higher
luminosities in the past, when the outflow was generated. If this is the case, the outflow
should start to disappear close to the central source in a timescale of decades,
as it moves away from the central source. In summary, we do not have a clear explanation 
for the large mechanical luminosity of the outflow and the apparent low bolometric luminosity
of the source at its center.

In Figure 1 we can also observe that the CO J=2$\rightarrow$1 redshifted 
integrated emission is stronger and wider but becomes detectable at a greater distance 
($\sim$ 5$''$) as compared to the blueshifted component. This suggests that the excitation 
of the CO molecule on opposite sides of the outflow
 is different, possibly due to anisotropic ambient cloud conditions. 
More observations at high angular resolution  with different molecular probes 
sensitive to different cloud properties will be quite important for studying
 the higher excitation material closely associated with the driving source.

Figures 3 show the channel maps of the CO J=2$\rightarrow$1 
emission for the red and blueshifted components, respectively. 
The greatest radial velocities are quite symmetric, reaching up to $\pm$80 km s$^{-1}$
from the ambient cloud velocity which is taken to be 5 km s$^{-1}$. The range of the outflow 
velocities, the central position of the outflow, and the orientation 
are in good agreement with the single-dish results of Rodr\'\i guez-Franco et al. (1999b).
On the redshifted side, there is a tendency for the gas to be located further 
from the exciting source with increasing velocity. 
However, the highest velocities are also found closer to the center. 
Furthermore, it is clear that for the CO redshifted emission, the higher velocities trace
 an extremely well collimated jet-like structure, while the shifted lower velocities 
 trace a more extended envelope that is consistent with a cylindrical shell, limb-brightened 
on the outer surface (see figure 4). These tendencies are less obvious at 
the blueshifted velocities possibly because of their fainter emission. 

The spatial structure of the high and low velocity gas of the bipolar outflow as a function 
of the radial velocities (Position-Velocity Diagram) along the major axis is shown 
in figure 5. The blue and redshifted components show that the
CO emission covers the same velocity range ($\pm$ 50-90 km s$^{-1}$) along the entire jet.
In the blueshifted component, the acceleration to the terminal velocity 
seems to be taken place immediately upon ejection (in a region of $\leq 5''$). 
A complete discussion on the possible ejection mechanisms
of the molecular outflows is given in  Bachiller (1996) and Lee at al. (2000).  
This behavior has also been observed in the molecular outflow 
IRAS 04166+2706 (Tafalla et al. 2004).
For the red component, there is also evidence that the molecular gas is decelerated   
along the jet, probably via interaction with the ambient gas.

\acknowledgments
We thank C. R. O'Dell for comments on an earlier version of this letter.   
We also thank all the SMA staff members for making these observations possible.
 LAZ acknowledges the Smithsonian Astrophysical Observatory 
for a predoctoral fellowship. LFR acknowledges the support
of DGAPA, UNAM, and of CONACyT (M\'exico).

\bibliographystyle{apj}

\clearpage

\begin{deluxetable}{l c c c c c c}
\small
\tablecaption{Outflow Parameters}
\tablehead{
\colhead{}      &
\colhead{Masa}                      &                       
\colhead{Momentum}                      &                            
\colhead{Energy}                        &
\colhead{Mechanical Luminosity}                    \\
\colhead{Component}      &
\colhead{[M$_{\odot}$]} &
\colhead{[M$_{\odot}$ km s$^{-1}$]} &
\colhead{[10$^{46}$ erg]}  &                    
\colhead{[L$_{\odot}$]}    & 
}
\startdata
Blueshifted & 0.20 &  16 & 1.3 & 170\\
Redshifted  & 0.52 &  42 & 3.3 & 440\\
Total       & 0.72 &  58 & 4.6 & 610\\
\enddata
\tablecomments{\footnotesize To calculate the energy and momentum we use a velocity of 
                             80 km s$^{-1}$. }
\end{deluxetable}

\begin{figure}
\centering
\includegraphics[scale=.75]{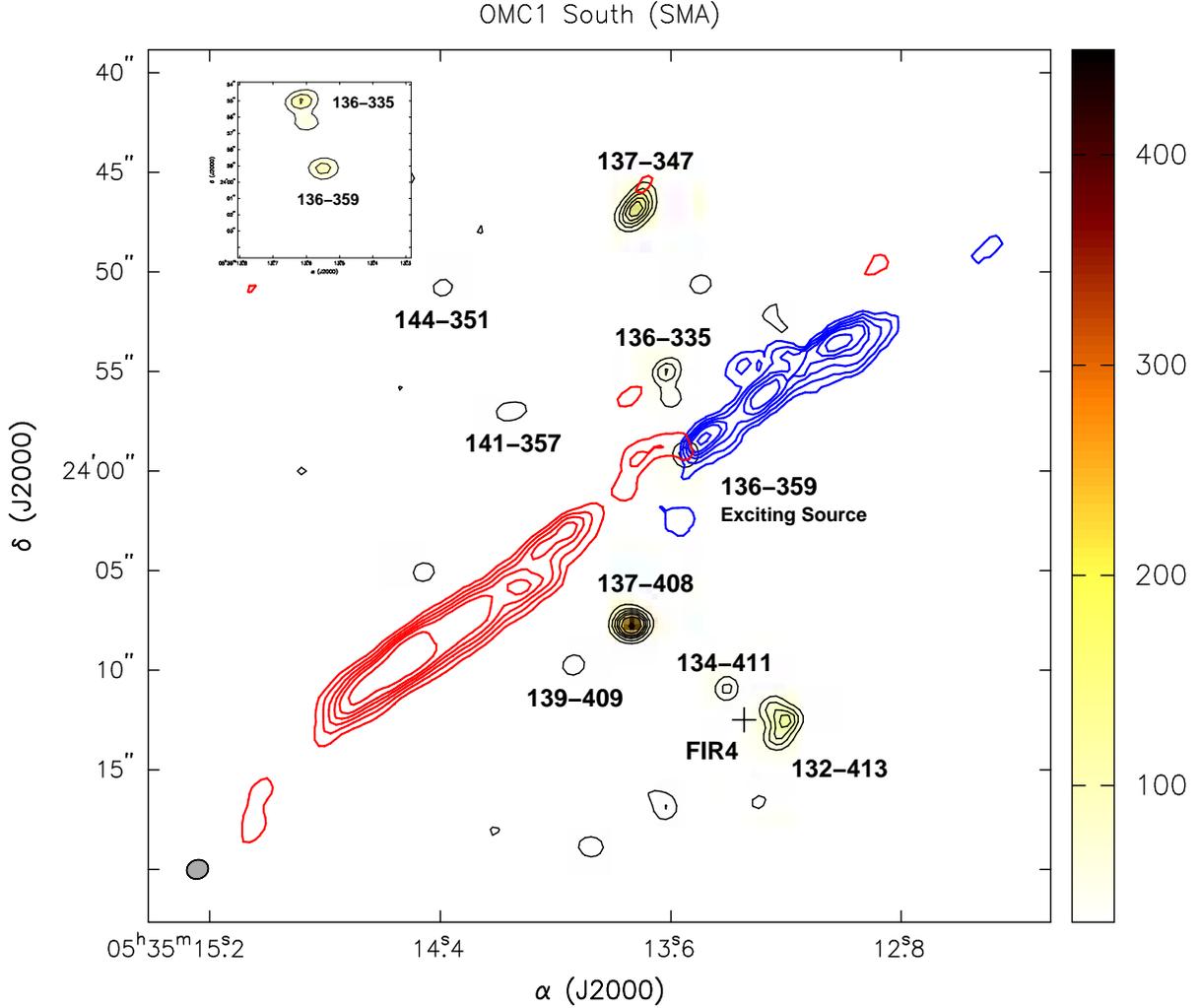}
\caption{\small SMA CO J=2$\rightarrow$1 moment zero (blue and red contours) and 
1.3mm  continuum image (black contours and yellow scale) towards the OMC1S region. The continuum 
contours are  3, 6, 9, 12, 15, 18, 20, and 30 times 11 mJy beam$^{-1}$,
the rms noise of the continuum image. The half power contour of the 
synthesized beam is shown in the bottom left. The CO contours are 3, 6, 9, 12, 
15, 18, 20 and 30 times 145 mJy beam$^{-1}$ km s$^{-1}$, the rms noise of the moment zero map.
The integration is over the velocity ranges: -80 to -26 km s$^{-1}$ (blue) and 
22 to 82 km s$^{-1}$ (red). The emission at ambient velocities (-25 to 21 km s$^{-1}$) 
is clearly extended and poorly sampled with the SMA, and is suppressed in this 
moment zero map. The cross denotes the position of the source 
FIR 4 (Mezger et al. 1990). The 3-$\sigma$ objects 139-409, 141-357 and 144-351 are 
likely real 1.3 mm continuum sources because they have 7 mm counterparts. The presumed exciting source 
is also shown at the center of the upper left box.}
\label{fig1}
\end{figure}

\clearpage

\begin{figure}
\centering
\includegraphics[scale=.65]{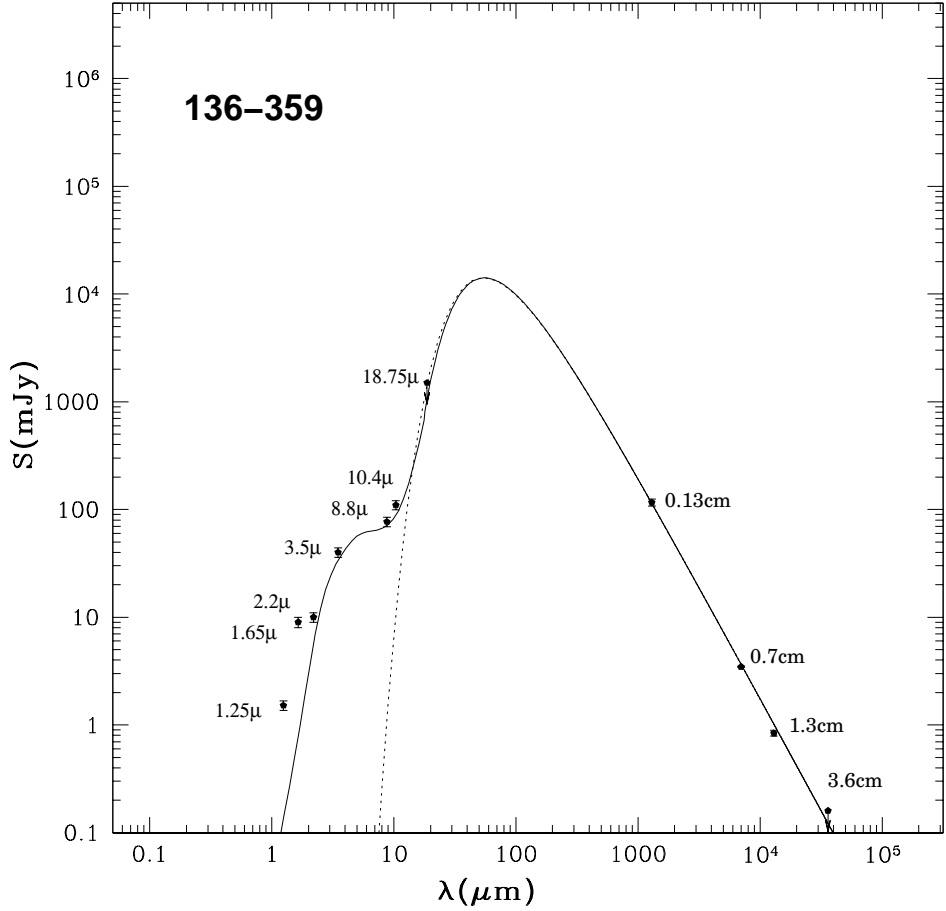}
\caption{\small Spectral energy distribution (SED) of 136-359 combining data obtained with the VLA (3.6 cm, 
Zapata et al 2004a; 1.3 cm , Zapata et al. 2004b; 0.7 cm, Zapata et al. 2005, in prep.), the Submillimer Array 
(1.3 mm, Zapata et al. 2005, in prep.), the Gemini South and Keck Observatories (8.8 and 18.75 $\mu$m,
 Smith et al. 2004), the UKIRT telescope (10.4 $\mu$m, Robberto et al. 2005), the FLW Observatory (3.5 $\mu$m, 
 Lada et al. 2000 and Lada et al. 2004), the USNO-IRCAM, (2.2 $\mu$m, 1.65 $\mu$m and 1.25 $\mu$m, 
 Gaume et al. 1998). The continuum curve is a fit to the spectrum using two modified blackbody 
 functions with different temperatures. The dotted curve is a fit for the cooler temperature component using
 a temperature of 83 K and $\beta$=0.13.} 
\label{fig2}
\end{figure}

\clearpage

\begin{figure}
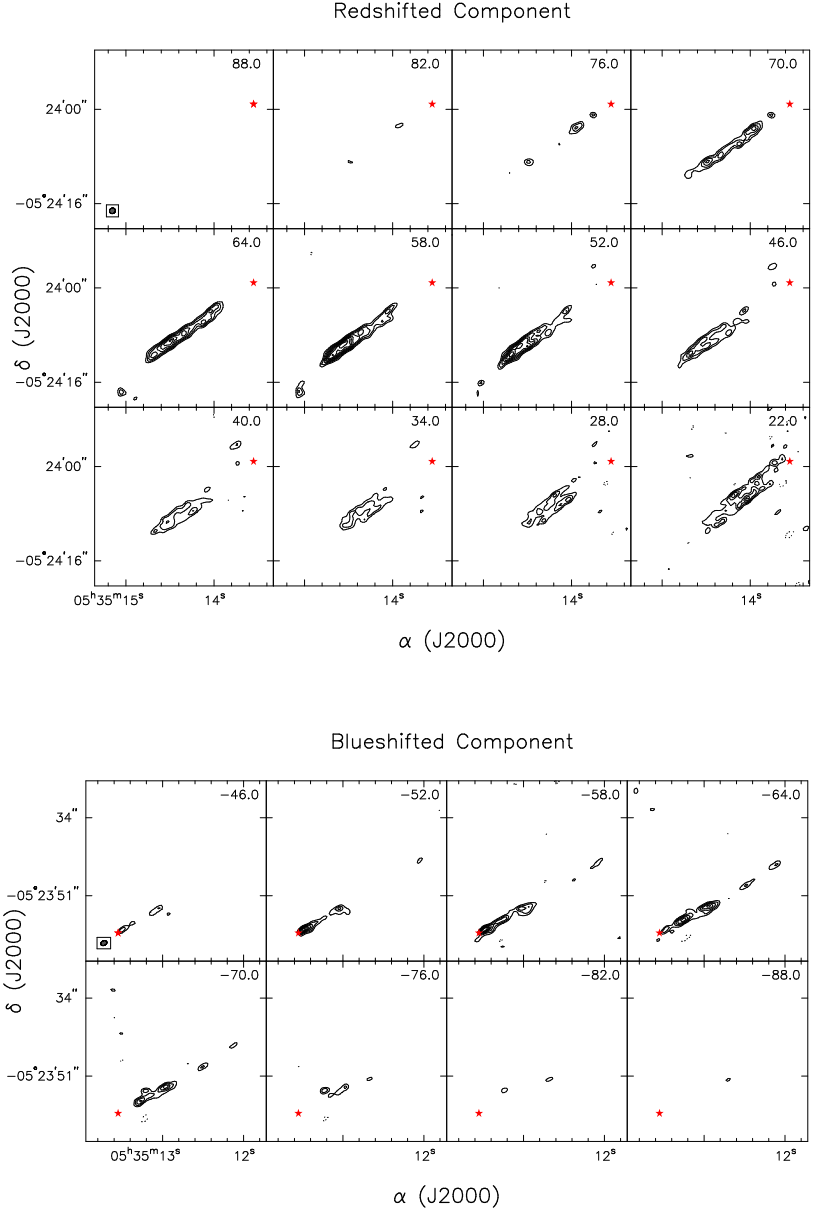

\begin{center}
\includegraphics[scale=.45, angle=-90]{f3.eps}\vspace{1cm}
\hspace{1cm}\includegraphics[scale=.45, angle=-90]{f4.eps}
\bigskip 
\caption{\small Channel maps of the CO J=2$\rightarrow$1 redshifted ({\bf upper}) and blueshifted ({\bf lower}) 
emission of the molecular outflow.
The emission is summed in velocity bins of 6 km s$^{-1}$. The central velocity is indicated in the upper 
right corner of each panel (the systemic velocity of the ambient
molecular cloud is about 5 km s$^{-1}$).  
The full width at half maximum of the synthesized beam is shown 
in the left bottom corner of the first box. 
The emission at ambient velocities (-21 to 21 km s$^{-1}$) 
is clearly extended and poorly sampled with the SMA, and is suppressed in these maps.
The contours are -3,-2, 2, 3, 4, 5, 6, 7, 8, 9, 10, 11, 12 times 0.12 Jy beam$^{-1}$, the rms noise 
of the image. The star denotes the position of the presumed exciting source.  }
\label{fig3}
\end{center}
\end{figure}

\clearpage

\begin{figure}
\centering
\includegraphics[scale=.75,angle=-90]{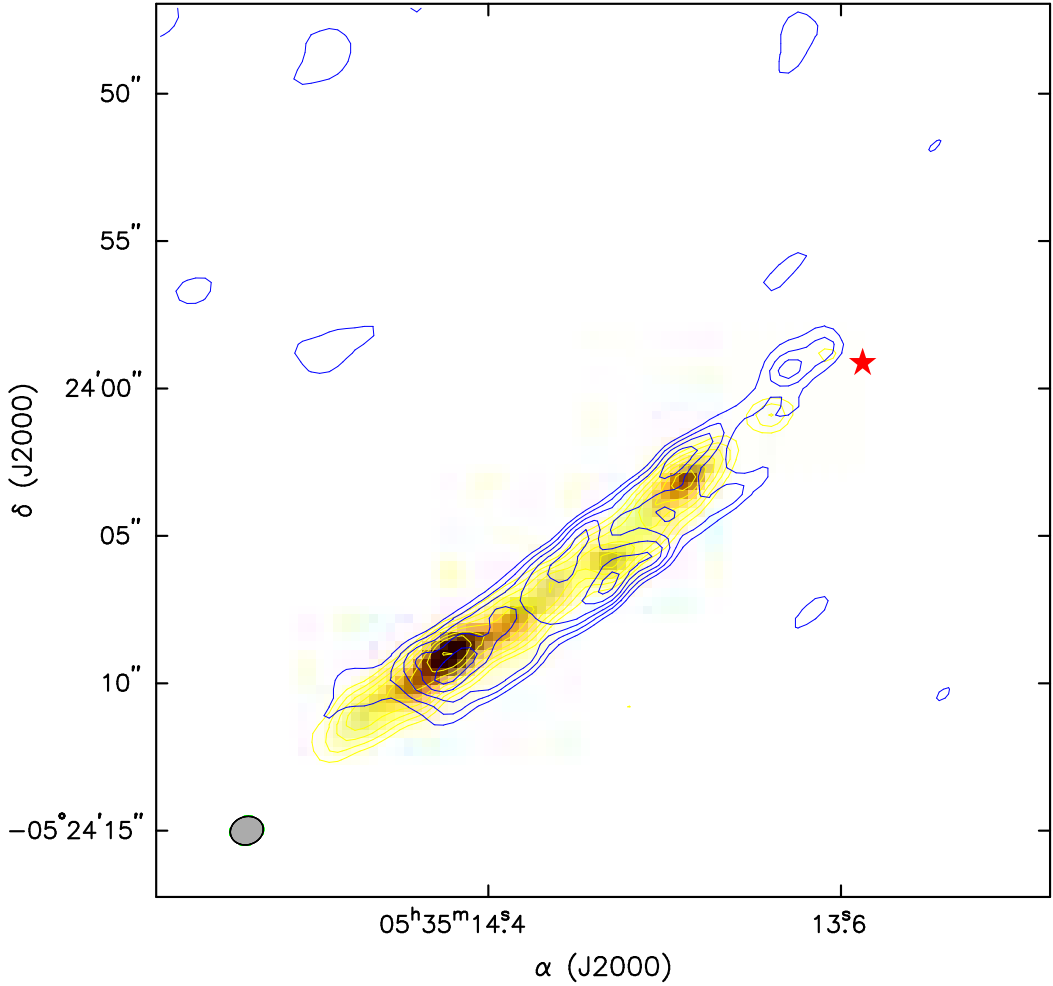}
\caption{\small SMA CO J=2$\rightarrow$1 moment zero of the high velocity outflow redshifted emission. 
The blue contours show the emission integrated over the velocity range: 18 to 40 km s$^{-1}$
 (low velocities). The contours are 2, 3, 4, 5, 6, and 7 times 195 mJy beam$^{-1}$ km s$^{-1}$,
 the rms noise of the image. 
The yellow colored image and contours show the emission integrated 
over the velocity range: 40 to 80 km s$^{-1}$ (high velocities).
The contours are 2, 3, 4, 5, 6, 7, 8, 10, 12, 15, 18 and 20 times 195 mJy beam$^{-1}$ km s$^{-1}$, 
the rms noise of the image. 
The full width at half maximum of the synthesized beam is shown in the left bottom 
corner. The red star denotes the position of the presumed exciting source.} 
\label{fig5}
\end{figure}

\clearpage

\begin{figure}
\centering
\includegraphics[scale=.55]{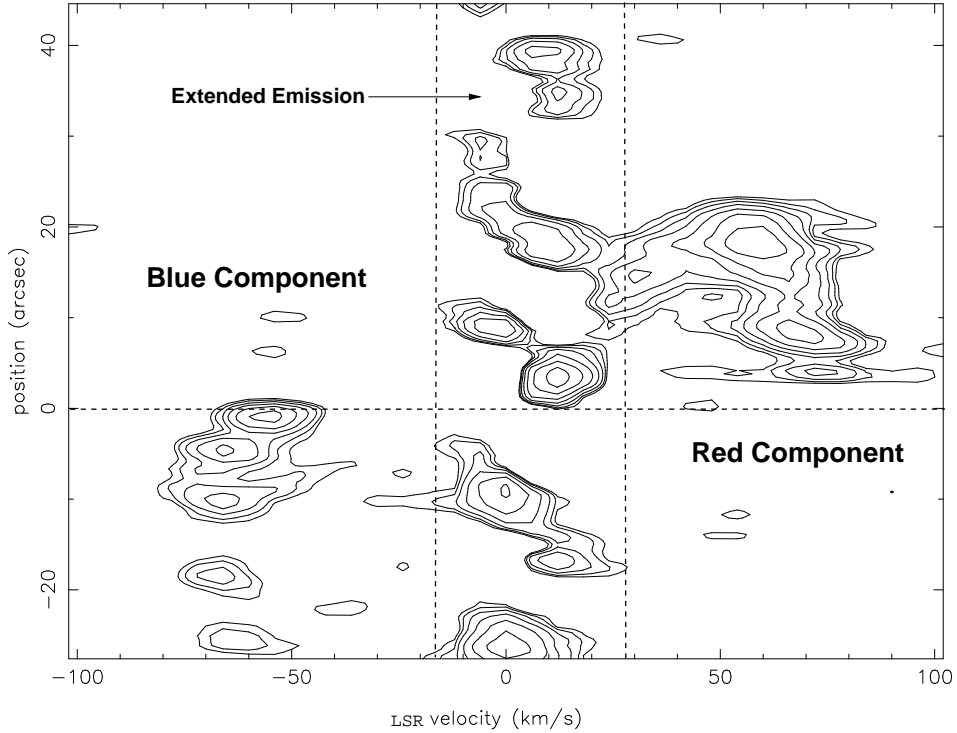}
\caption{\small Position-velocity diagram of the CO J=2$\rightarrow$1 emission, 
computed along the jet axis. The data set was smoothed in velocity and space. 
The lowest level corresponds to 0.1 Jy and the step is logarithmic
(0.06, 0.09, 0.15, 0.24, 0.38, 0.60, 0.95, 1.5, 2.4 and 3.8). The velocity and angular resolutions
are 6 km s$^{-1}$ and $\sim$ 1$''$, respectivity.
The vertical dashed lines bracket the ambient velocities that the interferometer samples poorly.
The horizontal dashed line indicates the position of the presumed exciting source.
We can observe how gas is decelerated along the redshifted component by comparing the 
position and velocity of the bullets in this component. The nearest bullet shows 
a high velocity (80 km/s), while the farthest bullet shows a low velocity (60 km/s). }
\label{fig6}
\end{figure}

\end{document}